\documentclass[printer]{aa} \usepackage[varg]{txfonts}
\usepackage{graphicx} \usepackage{color} \usepackage{amsbsy,amsmath}
\usepackage{amstext} \usepackage{natbib}
\usepackage{endnotes}
\begin{document}

\title{Planetary migration and the origin of the 2:1 and 3:2 (near)-resonant 
population of close-in exoplanets}
\titlerunning{Planetary migration and (near)-resonant systems}
\author{X.S. Ramos
        \inst{1},
        C. Charalambous
        \inst{1}, 
        P. Ben\'{\i}tez-Llambay
        \inst{1}$^{,}$\inst{2} 
        \and 
        C. Beaug\'e\inst{1}
        }
\authorrunning{X.S. Ramos et al.}
\institute{Instituto de Astronom\'\i a Te\'orica y Experimental, Observatorio 
        Astron\'omico, Universidad Nacional de C\'ordoba, Laprida 854, X5000BGR, 
        C\'ordoba, Argentina \\
        \email{xramos@oac.unc.edu.ar}
        \and
        Niels Bohr International Academy, Niels Bohr Institute, Blegdamsvej 17, 
        DK-2100 Copenhagen \O{}, Denmark \\
        }
\abstract{

We present an analytical and numerical study of the orbital migration
and resonance capture of fictitious two-planet systems with masses in
the super-Earth range undergoing Type-I migration. We find that,
depending on the flare index and proximity to the central star, the
average value of the period ratio, $P_2/P_1$, between both planets may
show a significant deviation with respect to the nominal value. For
planets trapped in the 2:1 commensurability, offsets may reach values
on\ the order of $0.1$ for orbital periods on the order of $1$ day,
while systems in the 3:2 mean-motion resonance (MMR) show much smaller
offsets for all values of the semimajor axis.

These properties are in good agreement with the observed distribution
of near-resonant exoplanets, independent of their detection method. We
show that 2:1-resonant systems far from the star, such as HD82943 and
HR8799, are characterized by very small resonant offsets, while higher
values are typical of systems discovered by Kepler with orbital
periods approximately a few days. Conversely, planetary systems in
the vicinity of the 3:2 MMR show little offset with no significant
dependence on the orbital distance.

In conclusion, our results indicate that the distribution of Kepler
planetary systems around the 2:1 and 3:2 MMR are consistent with
resonant configurations obtained as a consequence of a smooth
migration in a laminar flared disk, and no external forces are
required to induce the observed offset or its dependence with the
commensurability or orbital distance from the star.  }
\maketitle

\section{Introduction}

From the distribution of orbital period ratios of {\it Kepler}
multiplanet systems, two features stand out. On the one hand,
mean-motion resonances (MMRs) seem scarce and most of the systems
appear nonresonant. This may either indicate that planets formed in
situ (e.g., \citealp{2012ApJ...751..158H, 2013ApJ...770...24P,
  2014ApJ...780...53C}) or that planetary migration did not necessary
result in resonance trapping (e.g., \citealp{2012MNRAS.427L..21R,
  2013ApJ...778....7B}). A second characteristic is the existence of a
statistically significant number of planetary pairs close to
resonances (particularly the 2:1 and 3:2 MMRs), but with orbital
period ratios that are higher than the nominal value. Although such an
offset is expected from resonant dynamics, particularly for low
eccentricities, the observed values are much higher than obtained from
simulations of resonance trapping. It is generally believed that these
systems are near-resonant and located outside the libration domain.

Let us assume two planets of masses $m_1$ and $m_2$ orbiting a star
$m_0$, orbital periods $P_1 < P_2$ , and in the vicinity of a
first-order $(p+1)/p$ MMR.  We can then define the so-called {\it
  resonance offset} as
\begin{equation}
\Delta_{(p+1)/p} = \frac{P_2}{P_1} - \frac{(p+1)}{p} ,
\label{eq1}
\end{equation}
whose value indicates the distance from exact resonance. Although the
expected value depends on the planetary masses and eccentricities,
N-body and hydrodynamical simulations of resonant capture in the 2:1
MMR generally lead to $\Delta_{2/1} \simeq 10^{-3}$ (e.g.,
\citealp{2015MNRAS.453.4089S}) and similar values for the 3:2
resonance. However, a significant proportion of {\it Kepler} systems
exhibit much higher values (i.e., $\Delta_{2/1} \sim 0.04$ and
$\Delta_{3/2}\sim0.02$, see Figure \ref{fig1}), apparently placing
them outside the resonance domain.

The origin of these near-resonant systems is a dilemma. Divergent
migration through tidal evolution (e.g., \citealp{2012ApJ...756L..11L,
  2013AJ....145....1B, 2014A&A...570L...7D}) can reproduce some of the
dynamical properties of these systems, although the magnitude of the
tidal parameters required in many cases are not consistent with
current models (\citealp{2013ApJ...774...52L}).  Planetary migration
in a turbulent disk (\citealp{2013MNRAS.434.3018P}) can, under certain
circumstances, lead to near-resonant configurations, although it is
not clear whether this mechanism can in fact reproduce the observed
near-resonant distribution (e.g., \citealp{2013MNRAS.435.2256Q}).

More recently, \cite{2015MNRAS.453.1632M, 2016MNRAS.458.2051M} studied
the effect of more complex disk models on migration of planetary
pairs, including photo-evaporation and different opacity laws. They
found that several regions of divergent migration appear in the disk
that may lead to departure from exact resonance and even ejection from
the commensurabilities. However, they found little correlation between
their model and the observed distribution of bodies.

Perhaps a more basic question is how do we know these systems are
actually nonresonant. Calculations of the libration width require
information of the planetary masses, a luxury that is usually
unavailable in {\it Kepler} systems.  Estimations of the
eccentricities are also important since the libration widths are a
strong function of this element. However, the observed resonance
offset is believed to be larger than the libration domain, especially
for super-Earths and Neptune-size bodies.

In this work we present a simple analytical and N-body study of the
dynamics of two-planet systems leading to capture in the 2:1 and 3:2
MMRs in a laminar disk.  We show that the observed distribution of
resonance offsets amongst the exoplanetary population is consistent
with disk-planet interactions in a flared disk. Evidence from our
simulations seems to indicate that the main difference between the
warm and hot planetary systems is their distance from the central
star, and many of the {\it Kepler} near-resonant planets may actually
be inside the libration domain of the 2:1 and 3:2 commensurabilities.

The paper is structured as follows. In Section 2 we review the
distribution of all confirmed planets in the vicinity of the 2:1 and
3:2 MMR and analyze the observed offset as function of the orbital
period. Section 3 discusses how the resonance offset depends on the
planetary and disk parameters and how high values may be attained
without extreme eccentricity damping. We also analyze two
different analytical prescriptions for migration. Section 4 presents a
series of applications of our analytical model for flat and flared
disks and different mass ratios of the planets. In particular, we
search for values of the disk and planetary parameters that lead to
offsets similar to those observed in real systems. Applications to
several concrete cases are discussed in Section 5, while conclusions
close the paper in Section 6.

\section{Population near the 2:1 and 3:2 MMRs}

\begin{figure}[t!]
\includegraphics[width=0.45\textwidth,clip=true]{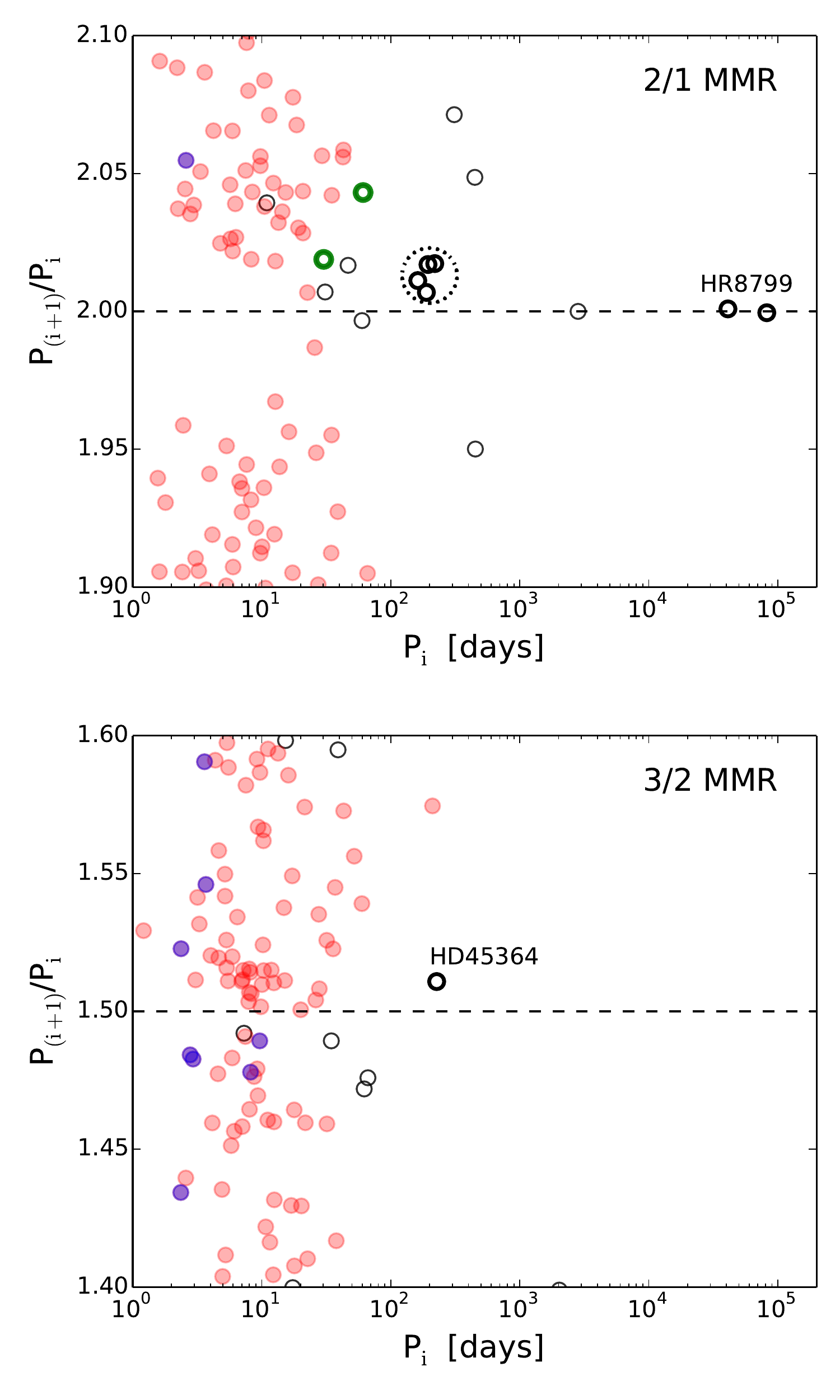}
\caption{Distribution of orbital period ratios of adjacent pair of
  planets as a function of the orbital period of the inner planet in
  the vicinity of the 2:1 MMR (top) and 3:2 MMR (bottom). Data
  obtained from {\tt exoplanet.eu}. The red circles identify confirmed
  planets detected by transits or TTV, while the blue circles
  correspond to recently validated {\it Kepler} planets proposed by
  \cite{2016ApJ...822...86M}. Planets discovered by any other method
  are depicted with open circles. In the top panel, open green circles
  denote the GJ876 system, while the dashed black circle groups the
  resonant (or near-resonant) members of HD73526, HD82943, HD155358,
  and HD92788.}
\label{fig1}
\end{figure}

Figure \ref{fig1} shows the distribution of orbital period ratios
$P_{\rm (i+1)}/P_{\rm i}$ for adjacent planets in multiplanetary
systems, as function of the orbital period $P_{\rm i}$ of the inner
member of each pair. The top plot shows the vicinity of the 2:1 MMR
while the bottom panel presents results for the 3:2
commensurability. In both cases, red circles correspond to planetary
systems detected by transits or transit time variations (TTV) and are
mostly fruits of the {\it Kepler} mission. The recently validated
systems proposed by \cite{2016ApJ...822...86M} are indicated
separately in blue, since they are not actually confirmed planets and
still subject to uncertainties. Finally, bodies discovered by all
other methods, particularly with radial velocity, are identified by
open black circles.

Both distributions show distinct features. Among the ``non-transit''
population, the 2:1 MMR hosts several well-known examples of resonant
pairs and even multiple resonances. The HR8799 planets are located
very far from the central star (orbital periods in excess of $10^4$
days) and appear to be locked in a Laplace-type resonance. The orbital
period ratio of the outermost planets are $2.0009$ and $1.9995$,
respectively, placing them extremely close to the location of exact
resonance (e.g., \citealp{2014MNRAS.440.3140G}.)

Closer to the star, and encompassed by an open dotted circle, lies a
group of planetary systems very close to the 2:1 resonance and
detected through radial velocity surveys. Perhaps the best known
examples are HD82943, whose two confirmed planets have a period ratio
of $\sim 2.017$ (\citealp{2014MNRAS.439..673B}), and HD73526 with a
value of $2.006$ (\citealp{2014ApJ...780..140W}). Uncertainties in the
orbital fits make it difficult to establish whether all other cases
are truly resonant (e.g., HD128311; see \citealp{2014ApJ...795...41M,
  2015MNRAS.448L..58R}), although stability considerations usually
require libration of the resonant angles.

The last example of RV systems in the vicinity of the 2:1 MMR is GJ876
with three planets locked in a Laplace resonance (e.g.,
\citealp{2010ApJ...719..890R, 2011CeMDA.111..235B}). Both adjacent
pairs are identified in the upper panel by broad open green
circles. Although the system also houses a much smaller inner planet, it
is located very close to the star and with negligible gravitational
interactions with the other members. The resonance offset for GJ876 is
larger than observed in other cases with orbital period ratios of
$\sim 2.02$ and $\sim 2.04$ (\citealp{2016MNRAS.455.2484N}).

The GJ876 system has been shown to be significantly chaotic
(\citealp{2013MNRAS.433..928M}), although stable for time spans
comparable to the age of the system. \cite{2015AJ....149..167B} argued
that the strong chaoticity is incompatible with a smooth migration in
a laminar disk, proposing instead that resonance capture occurred in a
turbulent disk characterized by low-amplitude stochastic
perturbations. However, it is unclear why turbulence would have
affected this system and not the other members of the resonant
population. On the other hand, \cite{2016MNRAS.460.1094M} showed that
the Laplace resonance for the GJ876 also contains an inner region of
very low chaoticity. Even though the current orbital fits place the
system in the more stochastic outer resonant domain, it is possible
that future observations may change this picture and be more
consistent with the regular inner zone.

The distribution of transit and TTV systems around the 2:1 MMR (Figure
\ref{fig1}, filled red circles) shows two main features that have been
the focus of several works over recent years. First, there is a
significant excess of planets with positive values of $\Delta_{2/1}$
with a median close to $0.04$ for $P_2/P_1 \in [2.0,2.1]$. Since the
planetary radii of these planets are consistent with masses in the
range of super-Earths and Neptunes, it is generally believed that
these systems are outside the librational domain and nonresonant in
nature. It is indeed curious to see that more distant systems, with
typically higher masses, have smaller offsets even though their
librational domain is significantly larger.

A second feature in the distribution of transit and TTV systems is a
significant increase in the value of $\Delta_{2/1}$ for planets closer
to the star.  \cite{2014A&A...570L...7D} also found a larger offset
closer to the central star by analyzing two subsets: the first with
$P_1 \le 15$ days and a second subset with $P_1 \ge 15$ days. These
authors attributed this dichotomy to tidal effects, which should be
important for orbital periods below (say) $P_1 \le 10-20$ days.
Figure \ref{fig1} shows similar results, but goes one step
further. The observed offset may indicate a possible smooth trend in
which $\Delta_{2/1} \sim 0.05$ for $P_1 \sim 1$ day, down to
$\Delta_{2/1} \sim 0.01$ for orbital periods $P_1 \sim 10^3$
days. Such a trend could suggest that the distribution of resonance
offsets is primarily dependent on the distance from the star and not
so much on the migration type. The case of GJ876 appears to comply
with this idea. Although detected through radial velocity and
consisting of giant planets that should have experienced Type-II
migration, their value of $\Delta_{2/1}$ is similar to that of {\it
  Kepler} systems and not to other, more distant, resonant cases
comprised of giant bodies.

The bottom panel of Figure \ref{fig1} shows results for the vicinity
of the 3:2 MMR. Presently there is only one confirmed resonant RV
system (i.e., ,HD45364; \citealp{2009A&A...496..521C}), with $P_1 \sim
226$ days and an offset of $\Delta_{3/2} \sim 0.01$. Thus, at least in
the case of radial velocity surveys, the resonant population of the
3:2 is much less abundant than observed in the 2:1.

The 3:2 transit and TTV systems also show significant differences with
respect to the 2:1 commensurability. On the one hand, the resonant
offset is noticeable smaller with a median of $\Delta_{3/2} \sim 0.01$,
as calculated for systems with $P_2/P_1 \in [1.5,1.6]$. This value is
on the same order as that observed for HD45364, indicating a shallower
dependence with the orbital period. In fact, as pointed out in
\cite{2014ApJ...795...85W}, the distribution of {\it Kepler} systems
is consistent with an accumulation close to exact resonance with small
offsets, and no significant gap is observed for low negative values of
$\Delta_{3/2}$. A second interesting feature is an apparent pile-up of
(near)-resonant systems with orbital periods $P_1 \sim 10$ days, which
is also not observed in the case of the 2:1 resonance.

In this paper we focus our attention on {\it Kepler} and other
close-in systems detected mainly through transits and TTV. Although in
most cases there is little information on the planetary masses, it is
believed they are relatively small bodies (Earth up to Neptune mass
range) and would have experienced planetary migration without having
carved a significant gap in the disk. We then use analytical models
for Type-I migration in an attempt to understand the observed
distribution and general trend.

\section{Resonance offset}

\subsection{Definition and analytical modeling}

In principle, there are two ways to explain the (near)-resonant
structure of the exoplanetary systems and its dependence on the
distance to the central star. In the first explanation, the planets
were initially trapped inside the resonance (exhibiting low values of
their offset) but later were driven out of the libration domain
through tidal effects. This appears reasonable for some systems in the
2:1 resonance, although it does not explain why the offset observed in
the 3:2 commensurability is much smaller. Moreover, tidal effects do not
seem to explain the trend observed in the 2:1 MMR for more distant
planets (e.g., Lee et al. 2013).

A second possibility is that resonance trapping, due to disk-planet
interactions, led to different values of $\Delta_{(p+1)/p}$ in
different parts of the disk. Thus, a planetary migration that was
halted far from the central star would be characterized by small
offsets, while this value increased from smaller orbital distances. This
is the mechanism that we analyze in the present work. However, before
simulating the migration process, it is important to review the
dynamical structure of planetary resonances and analyze the types of
motion expected for different values of $\Delta_{(p+1)/p}$.

\begin{figure}[t!]
\centering
\includegraphics[width=0.5\textwidth,clip=true]{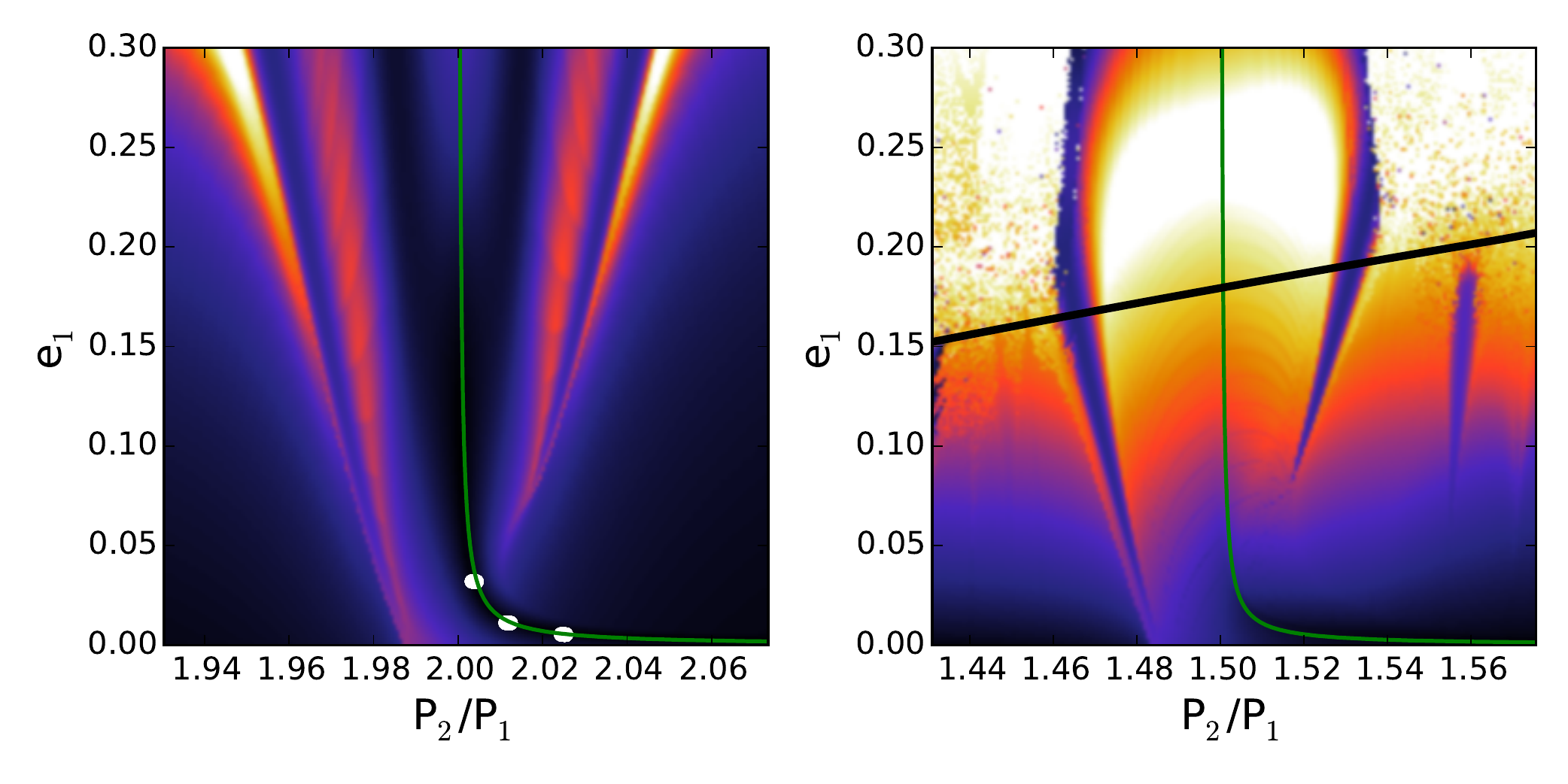}
\caption{Dynamical maps of $max(\Delta e)$ for the 2:1 (left) and 3:2
  MMR (right) for fictitious two-planet systems with $m_1 = 0.05
  m_{\rm Jup}$ and $m_2 = 0.10 m_{\rm Jup}$ orbiting a central star
  with mass $m_0 = 1 m_{\odot}$.  The orbit of the outer planet was
  initially circular with $a_2=1$ AU, and all angular variables where
  chosen equal to zero. Total integration time was $10^3$ years. The
  green lines indicate the location of the zero-amplitude ACR-type
  librational solutions, estimated from the simple analytical model
  (see Eq.  \ref{eq14}). The white dots in the left panel are the
  result of three N-body simulations of resonance trapping in the 2:1
  resonance. The broad black curve in the right plot corresponds to
  the Marchal-Bozis stability limit.  See text for details.}
\label{fig2}
\end{figure}

Figure \ref{fig2} shows two dynamical maps: the left panel for the 2:1
resonance and the right panel for the 3:2 commensurability. In each
case we integrated a series of two-planet systems with initial
conditions in a grid defined in the $(P_2/P_1,e_1)$ plane. The masses
of the planets were chosen equal to $m_1 = 0.05 m_{\rm Jup}$ and $m_2
= 0.10 m_{\rm Jup}$, and the central star assumed of solar mass. We
specifically chose a mass ratio $m_2/m_1 > 1$ to guarantee symmetric
fixed points for the resonant angles (e.g.,
\citealp{2006MNRAS.365.1160B, 2008MNRAS.391..215M}). The orbit of the
outer planet was initially considered circular with $a_2=1$ AU and all
angles equal to zero.

The color code corresponds to the maximum values of $|e_1(t) -
e_1(t=0)|$ (denoted here as $max(\Delta e)$) attained during a $10^3$
year integration time span. Darkened (lighter) tones are associated
with small (large) variations in the eccentricity of the inner
planet. For the 3:2 resonance, the broad black curve indicates the
stability limit as calculated using the Marchal-Bozis criterion
(\citealp{1982CeMec..26..311M}). Although the $max(\Delta e)$
indicator does not measure chaotic motion, it is an important tool
with which to probe the structure of resonances and help identify the
locus of stationary solutions, known as the ACR solutions (see
\citealp{2003ApJ...593.1124B, 2006MNRAS.365.1160B}), and the
separatrix delimiting the librational from the circulation domains
(e.g., \citealp{2015CeMDA.123..453R}).

In the same figure, the green lines show the approximate location of
the family of zero-amplitude ACR solutions
(\citealp{2003ApJ...593.1124B, 2006MNRAS.365.1160B}), which are
characterized by the simultaneous libration of both resonant angles
\begin{equation}
\begin{split}
\theta_1 &= (p+1) \lambda_2 - p\lambda_1 - \varpi_1 \\ 
\theta_2 &= (p+1) \lambda_2 - p\lambda_1 - \varpi_2 ,
\end{split}
\label{eq2}
\end{equation}
where $\lambda_i$ and $\varpi_i$ are the mean longitudes and
longitudes of pericenter of each planet. Using a simple Sessin-type
approximation for the resonant Hamiltonian (e.g.,
\citealp{1984CeMec..32..307S}), which neglects the secular
perturbations and is valid only for very low eccentricities, we can
estimate the value of the resonance offset as function of $e_1$, along
with a relationship between the eccentricities of both planets. These
give
\begin{equation}
\Delta_{(p+1)/p} = C_1 \; \frac{m_2}{m_0} \frac{1}{e_1} 
\hspace*{0.5cm} ; \hspace*{0.5cm} e_2 = C_2 \; \frac{m_1}{m_2} e_1
\label{eq3}
\end{equation}
(see \citealp{2004ApJ...611..517L}), where the coefficient $C_i$
depends solely on the semimajor axes ratio $a_1/a_2$. For the MMRs
under consideration, they take the following values: $C_1 \simeq 1.5 ,
C_2 \simeq 0.29,$ for the 2:1 resonance; and $C_1 \simeq 1.2 , C_2
\simeq 1$ for the 3:2 case.

For given resonances and planetary masses, the value of the offset is
inversely proportional to the equilibrium eccentricities; thus, very
low values of $e_i$ are necessary to obtain a significant deviation
from an exact resonance.  As an example, reproducing a resonant offset
of {$\Delta_{2/1} \simeq 0.05$} would require the eccentricities to
remain as low as $e_1 \sim 10^{-3}-10^{-4}$ during the final stages of
planetary migration.

If the disk-driven planetary migration is sufficiently slow and
smooth, and if the orbits remain almost circular, we expect the
orbital evolution to follow the pericentric branch into the
librational domain and exhibit low-amplitude oscillations of the
resonant angles. In such an ideal scenario, the final eccentricities
and resonant offset $\Delta_{(p+1)/p}$ depends on the relative
strength between the eccentricity damping and orbital migration
timescales (\citealp{2002ApJ...567..596L,2006MNRAS.365.1160B}),
usually denoted by $\tau_{e_i}$ and $\tau_{a_i}$, respectively. Thus,
in the adiabatic limit, the final outcome of a resonance trapping
depends fundamentally on the ratios ${\cal K}_i =
\tau_{a_i}/\tau_{e_i}$, which we denote here as the {\it K-factors}.

\cite{2005MNRAS.363..153P} deduced a closed relation between the
equilibrium eccentricities and the orbital and damping timescales,
which, after some simple algebraic manipulations, acquires the form
\begin{equation}
{e_2^2 \, {\cal K}_2(1-D) + e_1^2 \, {\cal K}_1 B \left( 
\frac{\tau_{a_2}}{\tau_{a_1}} \right) = \frac{D}{2}  \left( 1 - 
\frac{\tau_{a_2}}{\tau_{a_1}} \right)}, \\
\label{eq4}
\end{equation}
where the new coefficients 
\begin{equation}
D = \frac{1}{(p+1)} \left( 1 + \frac{a_1}{a_2} \frac{m_2}{m_1} \right)^{-1} 
\;\; ; \\
B = \frac{m_1n_2a_2}{m_2n_1a_1}+D
\label{eq5}
\end{equation}
depend solely on the resonance and the mass ratio of the
planets. Using equations (\ref{eq3})-(\ref{eq4}), we finally obtain an
expression for the resonance offset in terms of the migration
timescales as
\begin{equation}
\Delta_{(p+1)/p}^2 =
\frac{2}{D} \left( C_1 \frac{m_2}{m_0} \right)^2 
\frac{ \left[(1-D){\cal K}_2 \left( C_2 \frac{m_1}{m_2}\right)^2 + B{\cal 
K}_1 \left( \frac{\tau_{a_2}}{\tau_{a_1}} \right) \right]}
{ 1 - \left( \frac{\tau_{a_2}}{\tau_{a_1}} \right)} .
\label{eq6}
\end{equation}
Since resonance trapping only occurs in cases of convergent migration
(i.e., $\tau_{a_1} > \tau_{a_2}$), the denominator is always positive
and free from singularities.

\subsection{Offset and the K-factors}

Assuming the constant ratio $m_1/m_2$, expression (\ref{eq6}) shows
that the value of the offset is a linear function of $m_2$ indicating
that, for fixed values of the migration timescales, we would expect
larger offsets for more massive planetary pairs. Since this does not
seem to be the case, then the observed values of $\Delta_{(p+1)/p}$,
particularly for close-in systems, must be dominated by the
characteristics of the disk-planet interactions.

To test our model defined by expressions (\ref{eq3})-(\ref{eq6}), we
can perform simple tests with an N-body integrator including an ad hoc
external acceleration (e.g., \citealp{2008A&A...482..677C}):
\begin{equation}
\ddot{\bf r}_i = -\frac{1}{\tau_{a_i}} \biggl[ \frac{{\bf v}_i}{2} + 2{\cal K}_i 
\frac{({\bf v}_i \cdot {\bf r}_i) {\bf r}_i}{r_i^2} \biggr] ,
\label{eq7}
\end{equation}
where ${\bf r}_i$ is the radial vector and ${\bf v}_i$ the velocity
vector of the planet. Setting the values of $\tau_{a_i}$ at certain
prefixed values, we can then vary the values of ${\cal K}_i$ and
analyze its effects on the resonance offset.

We performed three N-body simulations of planetary migration using
this simple recipe. We used the same masses and initial orbit of the
outer planet as used to construct the dynamical maps. The initial
value of $a_1$ was chosen outside the 2:1 MMR and all angles equal to
zero. The orbital decay timescales were $\tau_{a_1} = 10^5$ and
$\tau_{a_2} = 7 \times 10^4$ years to guarantee convergent and
adiabatic migration. The values of the eccentricity damping timescales
$\tau_{e_i}$ were chosen to give equal K-factors for both planets
(i.e., ${\cal K}_1={\cal K}_2={\cal K}$). Each run considered a
different value of the K-factor: ${\cal K}=10^2$, ${\cal K}=10^3$ and
${\cal K}=10^4$.

The initial conditions were integrated for $10^5$ years. We assumed
that the disk began dispersing at $T_{\rm disp}=7 \times 10^4$ years
and its surface density reached zero at $T_{\rm disp} = 8 \times 10^4$
years. This process was modeled by reducing the magnitude of the
dissipative force smoothly with a hyperbolic tangent function, and its
timescale was sufficiently slow to allow the equilibrium solution of
the system to adapt adiabatically. The last $10\%$ of the outputs were
used to calculate averaged values of the mean motions and
eccentricities. The white symbols in the left panel of Figure
\ref{fig2} show the final results.

The run with ${\cal K} = 10^2$ ended with $e_1 \simeq 0.03$ and
$P_2/P_1 \simeq 2.004$, that is, very close to exact resonance and
reminiscent of the individual pairs of HR8799. The second run, with
${\cal K} = 10^3$ yielded a lower equilibrium eccentricity ($e_1
\simeq 0.01$) and a slightly larger offset ($P_2/P_1 \simeq 2.01$),
similar to that of several of the resonant RV planets (e.g., HD82943).
Finally, the simulation with ${\cal K} = 10^4$ showed the lowest final
eccentricity and largest offset; i.e., $P_2/P_1 \simeq 2.03$, which is
on the order of that observed for GJ876. All these values are in
perfect agreement with the estimations deduced from expressions
(\ref{eq3}) and (\ref{eq6}).

K-factors on the order of $10^4$ are already much higher than
predicted by linear models of Type-I disk-planet migration (e.g.,
\citealp{2000MNRAS.315..823P, 2004ApJ...602..388T,
  2008A&A...482..677C}). From these works, in the low eccentricity
limit, we may write
\begin{equation}
\tau_{a_i} = Q_a \frac{t_{{\rm wave}_i}}{H_{r_i}^2} 
\hspace*{0.5cm}  ; \hspace*{0.5cm} 
\tau_{e_i} = Q_e \frac{t_{{\rm wave}_i}}{0.780}  , 
\label{eq8}
\end{equation}
where $H_{r_i}$ is the disk aspect-ratio in the position of each
planet. The value $Q_e$ is a constant and $Q_a = Q_a(\alpha)$ is a
function of the slope $\alpha$ of the surface density profile. The
linear model by \cite{2002ApJ...565.1257T} found $Q_a^{-1} \simeq 2.7
+ 1.1 \alpha$, while numerical fits with hydro-simulations usually
lead to slightly different functional forms (e.g.,
\citealp{2010ApJ...724..730D}). Finally,
\begin{equation}
t_{{\rm wave}_i} = \frac{m_0}{m_i} \frac{m_0}{\Sigma(a_i) a_i^2} 
\frac{H_{r_i}^4}{\Omega(a_i)} ,
\label{eq9}
\end{equation}
is the typical timescale for planetary migration. Here $\Omega(a_i) = 
\sqrt{{\cal G} (m_0 + m_i)/a_i^3}$ is the (circular) Keplerian angular velocity 
and ${\cal G}$ the gravitational constant. 

We assume a laminar disk whose surface density $\Sigma$ and the
aspect-ratio $H_r$ have a power-law dependence with the distance from
the star,
\begin{equation}
\Sigma(r) = \Sigma_0 r^{-\alpha} 
\hspace*{0.5cm} ; \hspace*{0.5cm} 
H_r(r) = H_0 r^{f}  , 
\label{eq10}
\end{equation}
where $\Sigma_0$ and $H_0$ are the values at $r=1$ AU. The exponent
$f$ is usually referred to as the disk-flare and is zero for flat
disks where the aspect-ratio is constant.

Finally, the quantity $Q_e$ in expression (\ref{eq8}) is an artificial
ad hoc constant parameter introduced by \cite{2006A&A...450..833C} to
fit the results of hydrodynamical simulations. They found that the
best results were obtained with $Q_e \sim 0.1$, a value that we also
adopt here.

From equations (\ref{eq8})-(\ref{eq10}) we can now obtain a simple
explicit form for the ${\cal K}$-factor as
\begin{equation}
{\cal K}_i \equiv \frac{\tau_{a_i}}{\tau_{e_i}} = \frac{0.78 Q_a}{H_0^2 Q_e} 
r_i^{-2f}.
\label{eq11}
\end{equation}
Assuming flat disks ($f=0$) and typical values of $H_0 \sim 0.05$, we
obtain factors on the order of ${\cal K}_i \sim 300$ for any distance
from the star, which are the values usually employed in N-body
simulations of planetary migration and resonance capture (e.g.,
\citealp{2013MNRAS.435.2256Q, 2015MNRAS.453.4089S}). A flared disk ($f
> 0$) not only leads to a dependence of ${\cal K}_i$ on the semimajor
axis, but also increases its value for close-in planets. Adopting $f =
0.25$ we obtain similar values as before for $a = 1$ AU but
significantly higher (${\cal K}_i \sim 10^3$) for $a = 0.05$ AU.

Although this road seems promising, the increase in the K-factor is
not sufficient. A resonance offset of $\Delta_{2/1} \simeq 0.05$, such
as those observed in close-in systems, would require ${\cal K}_i \sim
10^5$, which at least seems improbable. \cite{2015MNRAS.449.3043X}
proposed that planetary traps close to the central star could halt
planetary migration while not affecting the eccentricity damping
timescale, leading to an artificial increase of the corresponding
magnitude of ${\cal K}$. Even if such an idea seems possible, it would
not explain the observed trend of decreasing offset as function of the
semimajor axis, nor would it be expected to work for orbital periods
on the order of $10^2$ days.

\subsection{Goldreich \& Schlichting prescription}

\cite{2014AJ....147...32G} proposed to modify expressions (\ref{eq8})
to account for the contribution of eccentricity damping to changes in
the semimajor axis associated with (partial) conservation of the
angular momentum. Thus, the effective characteristic timescale for
orbital migration should actually be given by
\begin{equation}
{\frac{1}{\tau_{{a_{eff}}_i}} = \frac{1}{\tau_{a_i}} + 
2 \beta \frac{e_i^2}{\tau_{e_i}},}
\label{eq12}
\end{equation}
where $\tau_{a_i}$ and $\tau_{e_i}$ maintain the same form as
equations (\ref{eq8}) and $\beta$ is a factor that quantifies the
fraction of the orbital angular momentum preserved during the
migration. The value $\beta = 1$ if the orbital angular momentum is
completely preserved (e.g., tidal effects for synchronous orbits),
while $\beta < 1$ in other cases. \cite{2014AJ....147...32G} refer to
estimations by \cite{2004ApJ...602..388T} that suggest $\beta \simeq
0.3$ for Type-I disk-planet interactions.

The modified migration timescale, of course, also changes the value of
the K-factor, leading to a new ``effective'' form,
\begin{equation}
{\cal K}_{{eff}_i} \equiv \frac{\tau_{a_{{eff}_i}}}{\tau_{e_i}} = \frac{{\cal 
K}_i}{1+ 2 \beta e_i^2 {\cal K}_i},
\label{eq13}
\end{equation}
where ${\cal K}_i$ is given by equation (\ref{eq11}). Since the
equilibrium eccentricities within a resonance usually are on the order
of $e_i \sim 1/{\cal K}_i$, at least at the order of magnitude level,
we expect that the new value of the K-factor should not be much
different from its original. Moreover, since the denominator in
(\ref{eq13}) is always larger than unity, it should be verified that
${\cal K}_{{eff}_i} < {\cal K}_i$. Consequently, angular momentum
considerations in the migration prescription do not change the picture
significantly and cannot account for K-factors necessary to explain
the observed resonance offsets.

The change in the effective migration timescale also affects equation
(\ref{eq6}) for the resonance offset. After some algebraic
manipulations, we found a new version given by
\begin{equation}
\Delta_{(p+1)/p}^2 =
\frac{2}{D} \left( C_1 \frac{m_2}{m_0} \right)^2 
\frac{ \left[(1-D(1+\beta)){\cal K}_2 \left( C_2 \frac{m_1}{m_2}\right)^2 + 
(B+D\beta){\cal K}_1 \left( \frac{\tau_{a_2}}{\tau_{a_1}} \right) \right]}
{ 1 - \left( \frac{\tau_{a_2}}{\tau_{a_1}} \right)} 
\label{eq14}
,\end{equation} where the expressions for $\tau_{a_i}$ and ${\cal
  K}_i$ are those defined in (\ref{eq8}) and (\ref{eq11}).

\section{Differential migration and resonance offset}

Even if planetary migration occurs with limited values of the
K-factors, a significant value for the resonance offset may still be
obtained if the denominator in expression (\ref{eq14}) is sufficiently
small. As mentioned previously, this factor measures the differential
migration timescale between both planets. In order for resonance
capture to be possible, the outer planet must migrate faster than its
inner companion (i.e., $\tau_{a_2} < \tau_{a_1}$) leading to positive
values of $(1 - \tau_{a_2}/\tau_{a_1})$ and, consequently, real values
of $\Delta_{(p+1)/p}$.

From the expressions for $\tau_{a_i}$, the condition for convergent
migration implies that the mass ratio $m_2/m_1$ between the planets
must satisfy the condition
\begin{equation}
\frac{m_2}{m_1} \ge \left( \frac{m_2}{m_1} \right)_{\rm min} \equiv \biggl( 
\frac{P_2}{P_1} \biggr)^{\frac{2}{3}(2f+\alpha-0.5)} .
\label{eq17}
\end{equation}
For given initial conditions, this equation defines the limiting mass
ratio leading to capture in different commensurabilities. In the
general case of flared disks, the possibility of trapping in a given
resonance not only depends on the mass ratio itself but also on the
initial orbital separation between the planets.

Complying with condition (\ref{eq17}), the next question is how does
the value of the offset vary as a function of the mass ratio and
orbital distance.  Particularly, we wish to analyze whether there
exists any set of parameters that appear consistent with the
observational distribution of resonant and near-resonant systems.

\begin{figure*}[t!]
\centering
\includegraphics[width=0.98\textwidth,clip=true]{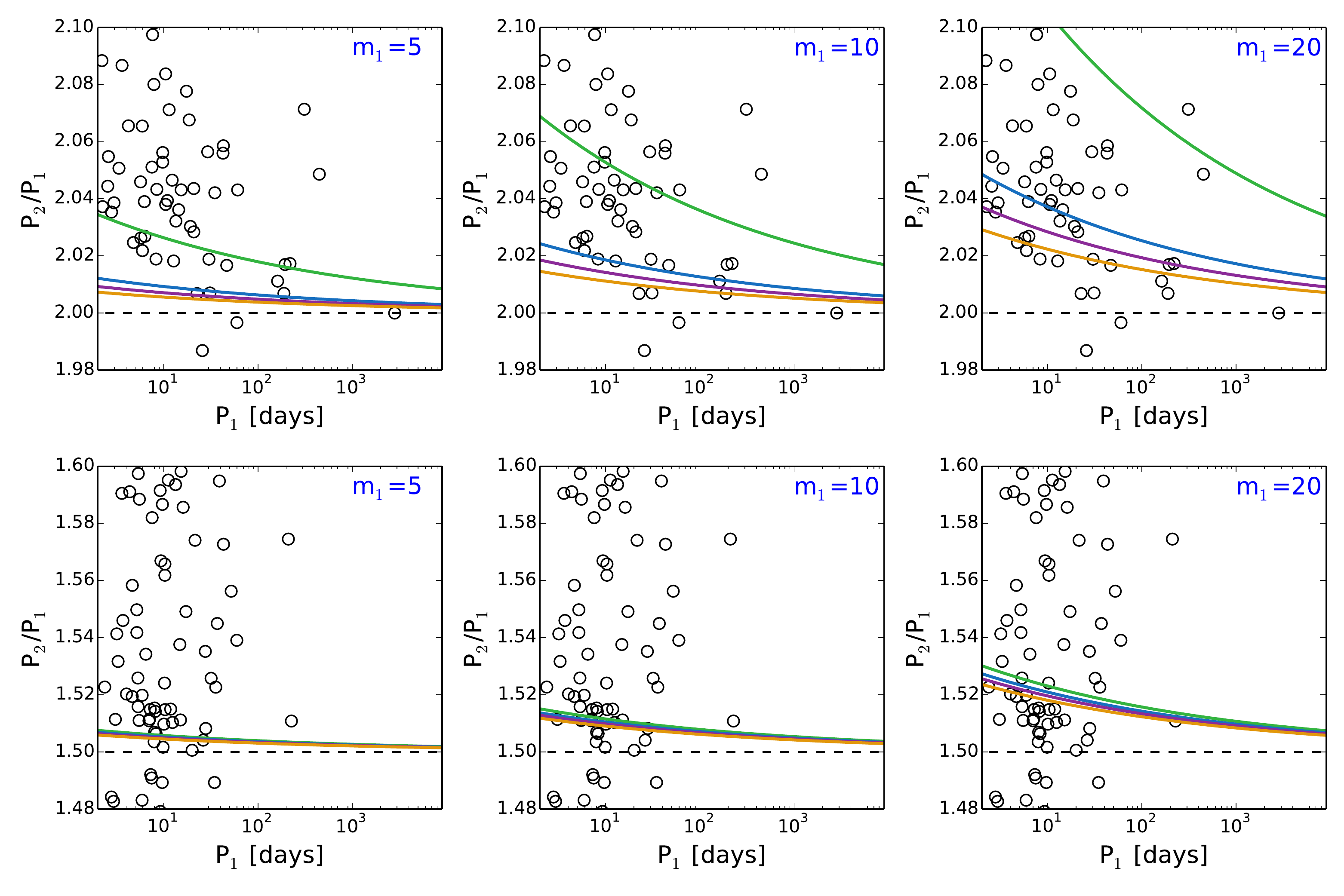} 
\caption{Orbital period ratio $(p+1)/p + \Delta_{(p+1)/p}$ as a
  function of the period of the inner planet, as predicted by
  expression (\ref{eq14}) for flared disks with $H_0 = 0.05$,
  $\alpha=1$, and $f=0.25$. Top panels represent the 2:1 MMR while
  bottom plots represent the 3:2 resonance. Each graph assumes a
  different value for the mass (in units of $m_\oplus$) of the inner
  planet, whose value is indicated in the top right-hand
  corner. Colored curves correspond to different mass ratios: $m_2/m_1
  = 1.6$ (green), $m_2/m_1 = 1.7$ (blue), $m_2/m_1 = 1.8$ (purple),
  and $m_2/m_1 = 2$ (orange). Open black circles denote the
  distribution of confirmed planets.}
\label{fig3}
\end{figure*}

Figure \ref{fig3} shows an application of expression (\ref{eq14}) for
$H_0 = 0.05$, $\alpha=1$ and $f=0.25$, in the form of the predicted
equilibrium values of the orbital period ratio $P_2/P_1 = (p+1)/p +
\Delta_{(p+1)/p}$ as a function of $P_1$. Results for the 2:1 MMR are
shown in the top panels, while those for the 3:2 are presented in the
bottom plots. Observational data is shown in open black circles. Each
panel corresponds to a different value of $m_1$, while each color
curve corresponds to a different value of the mass ratio $m_2/m_1$.

In all cases the offset increases with smaller semimajor axes, although
its magnitude is a strong function of both $m_1$ and $m_2/m_1$. As
expected from equation (\ref{eq17}), mass ratios close to
$(m_2/m_1)_{\rm min}$ show high values of $\Delta_{(p+1)/p}$, although
(also expected) its value increases linearly with $m_1$. For the 2:1
MMR, and for these values of the disk parameters, the observed
distribution of resonance offsets amongst exoplanetary pairs does not
seem compatible with inner masses much lower than $m_1 \sim 10
m_\oplus$.  Although this limiting case is marginally adequate, a much
larger diversity in values is obtained assuming $m_1 = 20 m_\oplus$,
where practically all of the observed offsets are encompassed by the
model, even if the mass ratios are not restricted to values very close
to unity.

The bottom plots, corresponding to the 3:2 MMR, show a different
story. In all cases the offsets increase with smaller distances to the
star, but show very little dependence with both $m_1$ and the mass
ratio. The best results are obtained assuming $m_1 \sim 10 m_\oplus$
(at least for these disk parameters). The lower dispersion in
$\Delta_{3/2}$, with respect to $\Delta_{2/1}$, is also in good
agreement with the observations.

\begin{figure}[t!]
\centering
\includegraphics[width=0.50\textwidth,clip=true]{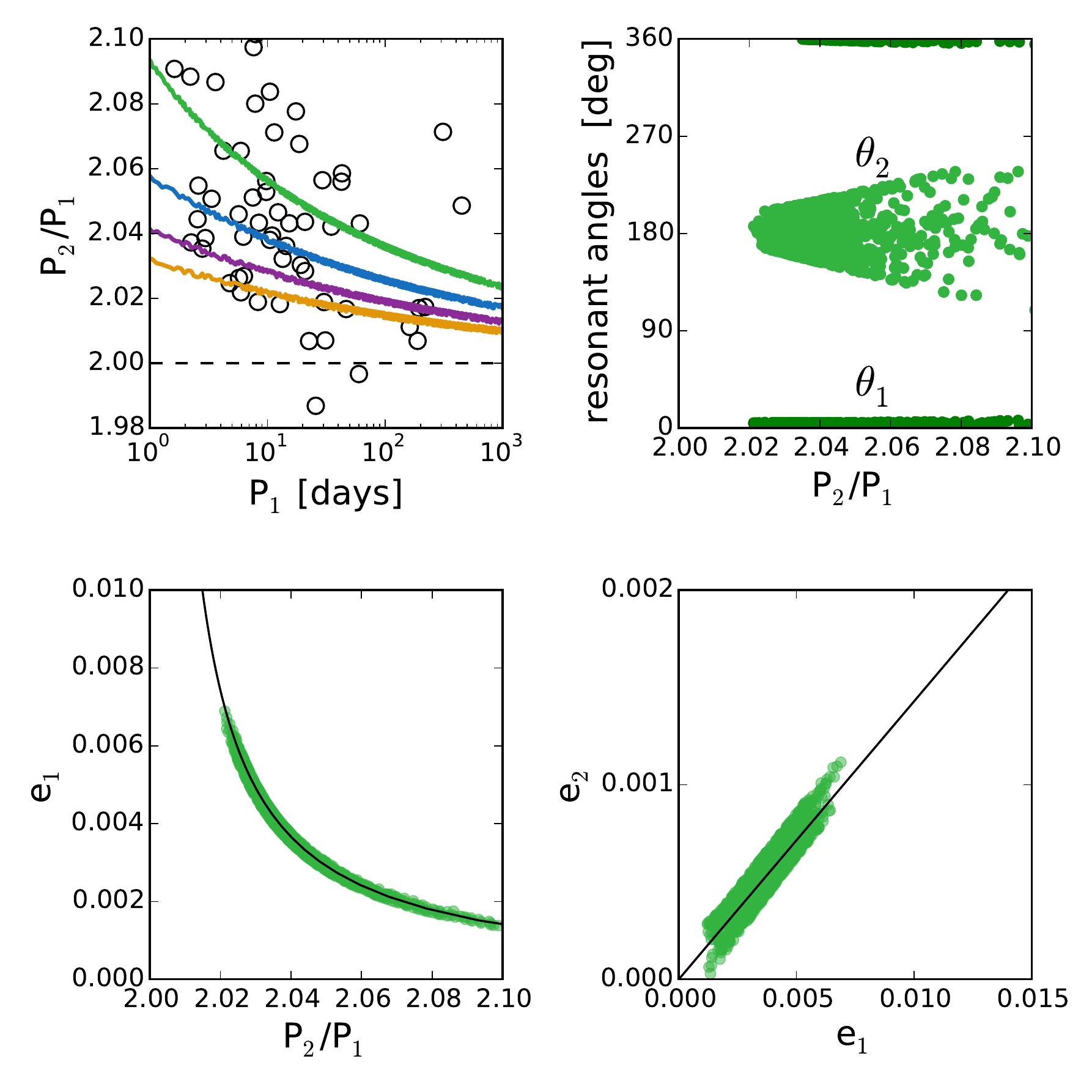} 
\caption{Top left panel shows the result of four N-body simulations
  leading to capture in the 2:1 MMR. In all cases $m_1=20
  m_\oplus$. Color lines indicate different mass ratios: $m_2/m_1=2$
  (orange), $m_2/m_1=1.8$ (purple), $m_2/m_1=1.7$ (blue), and
  $m_2/m_1=1.6$ (green). The remaining panels show different dynamical
  characteristics of the orbital evolution within the resonance of the
  run with $m_2/m_1=1.6$. The black continuous lines in the bottom
  panels represent the analytical predictions given by expressions
  (\ref{eq3}).}
\label{fig4}
\end{figure}

These results were obtained from our analytical model leading to
expression (\ref{eq14}) and still require numerical confirmation. The
upper left-hand plot of Figure \ref{fig4} show four N-body
simulations, where planetary migration was modeled according to
equation (\ref{eq7}), incorporating the Goldreich-Schlichting
prescription through expressions (\ref{eq12}) and (\ref{eq13}). We
assumed $\beta = 0.3$, $H_0 = 0.05,$ and a surface density of
$\Sigma_0 = 400$ gr/cm$^2$. The factors $Q_a$ and $Q_e$ were chosen as
described at the end of Section 3.2. In all cases we simulated capture
into the 2:1 MMR with $m_1=20 m_\oplus$ and different values for the
mass ratio $m_2/m_1$. As predicted, the offset effectively increases
closer to the star, showing a very good agreement with the
observations.

The other three plots show different aspects of the simulation with
$m_2/m_1 = 1.6$, where the value for the mass ratio leads to the
largest offset. The top right-hand graph shows the evolution of
both critical angles as function of $P_2/P_1$, where we have neglected
the data points prior to resonance lock. The main resonant angle
$\theta_1$ exhibits a low-amplitude libration around zero, even for
offsets $\Delta _{2/1} \simeq 0.1$, which are values that are usually
associated with nonresonant motion. The auxiliary resonant angle
$\theta_2$ also librates throughout the evolution of the system,
although with increasing amplitude. Thus, even in this extreme
numerical example, the two-planet system remains locked in an ACR-type
solution regardless of its proximity to (or separation from) exact
resonance.

The two bottom panels of Figure \ref{fig4} show complementary
information: the eccentricity of the inner planet as function of
$P_2/P_1$ (left) and the relation between both planetary
eccentricities (right). Thin black continuous lines indicate the
expected functional form as given by expressions (\ref{eq3}) obtained
for ACR solutions with the Sessin resonant Hamiltonian. The agreement
with the N-body run is excellent, indicating that the increase in
resonance offset follows closely the loci of low-amplitude ACR
solutions and the evolution of the system continues to be dominated by
the commensurability even for high values of $\Delta_{2/1}$.

We performed a large series of N-body simulations covering different
values of the planetary masses and flare index, observing the same
dynamical evolution as described in Figure \ref{fig4}. Thus, contrary
to expectations, large offsets may still be linked to resonant capture,
even if the resulting ACR solution corresponds to a kinematic (and not
dynamical) libration domain (e.g., \citealp{1983CeMec..30..197H}).

\section{Application to individual systems}

The mechanism proposed in this paper depends heavily on the individual
masses and mass ratio of the planets located in the vicinity of the
2:1 and 3:2 MMR.  The increase in the resonance offset close to the
star is most prominent when the mass ratio is close to unity and $m_1$
at least on the order of $10 m_\oplus$. More importantly, in order for
convergent migration to take place, the outer planet must be more
massive than its inner companion by a factor strongly dependent on the
flare index assumed for the disk.

It is not easy, however, to evaluate whether observed exoplanets in
(near)-resonant configurations validate this scenario. Most systems
where Type-I migration is applicable have been detected by transit,
making accurate measurements of the masses a rare commodity.

\begin{figure}[t!]
\centering
\includegraphics[width=0.50\textwidth,clip=true]{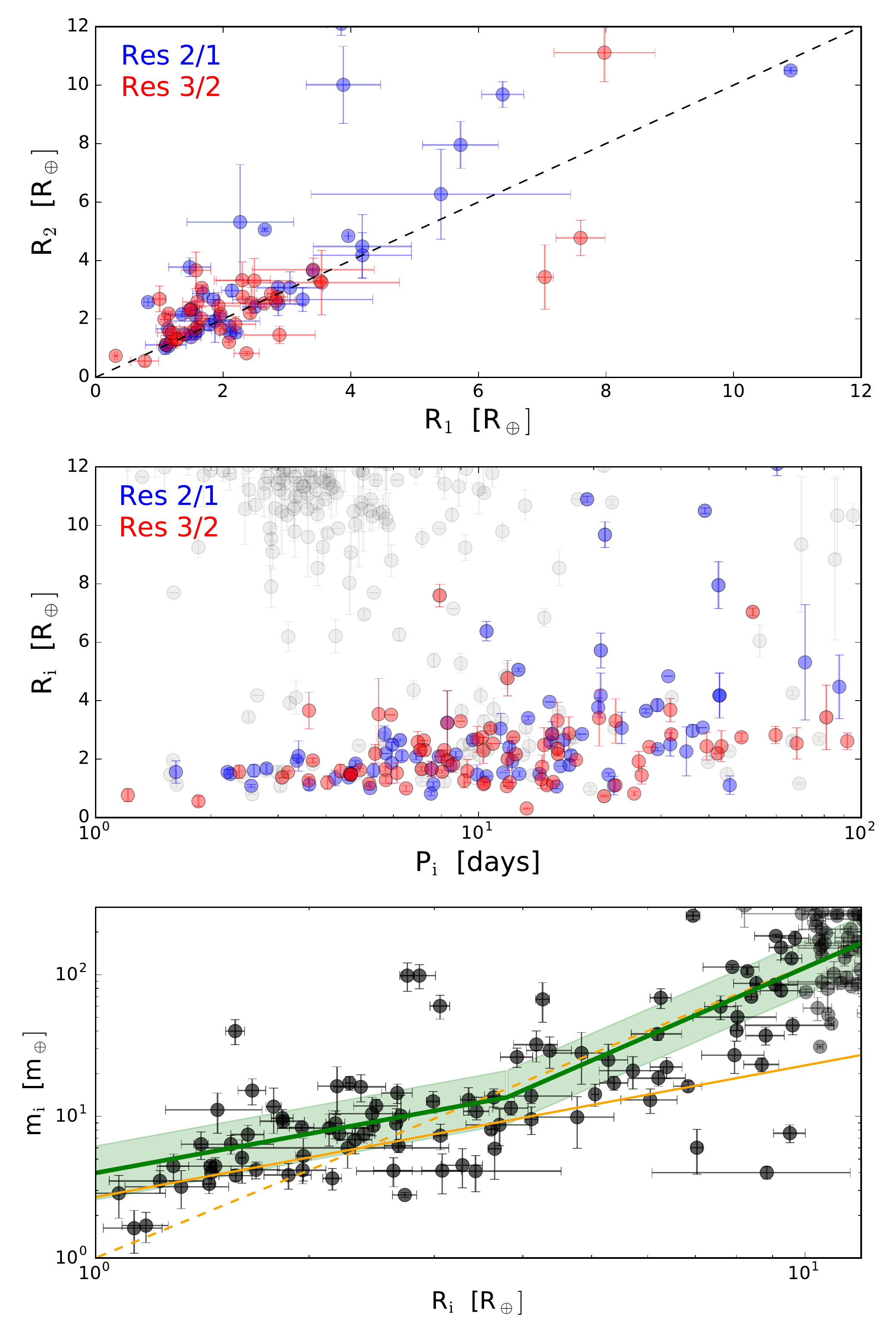} 
\caption{Top: Distribution of radii between confirmed planets in the
  vicinity of the 2:1 (blue) and 3:2 (red) MMR. Dashed line
  corresponds to equal-size bodies ($R_1=R_2$). Middle: Radius as a
  function of the orbital period. Light gray circles indicate the
  values of planets outside both resonances. Bottom: Planetary mass
  $m_i$ vs. radius $R_i$ for all planets (bodies larger than Jupiter
  shown in gray). Dashed orange line corresponds to the so-called
  $R^{2.06}$-law by Lissauer et al. (2011), while the fit proposed by
  Weiss \& Marcy (2014) is shown as a continuous orange line. Our own
  fit, which is calculated for all planets smaller than Jupiter, is
  presented as a thick green line. Values within $1\sigma_{\rm dis}$
  dispersion are contained within the light green region.}
\label{fig5}
\end{figure}

In order to have at least some statistical validation, we analyzed the
known exoplanetary population of confirmed planets, as published in
the {\tt exoplanet.eu} web site in mid-December 2016. Results are
shown in Figure \ref{fig5}. In the top panel we plotted the
distribution of radii ($R_2$ versus $R_1$) for all planetary pairs in
the vicinity of the 2:1 and 3:2 resonances.  We only considered those
bodies for which the uncertainties in radius were smaller than $40\%$
and resonance offsets in the interval were $(0,0.1]$. Most of the
  planetary pairs have radii below $4 R_\oplus$, although a few belong
  to the giant planet domain up to and beyond on Jupiter radius
  $R_{\rm Jup}$. Although in most cases $R_2/R_1 > 1$, a few are
  characterized by bigger inner planets.  However, without estimations
  of the planetary density, it is not possible to say whether the
  masses also comply with this condition and would not be possible
  candidates for resonance capture through inward planetary migration.

The middle plot shows the distribution of radii as function of the
orbital period.  The color code is the same as above, while all bodies
outside the 2:1 and 3:2 MMRs are indicated by light gray
circles. While the population of both resonances show decreasing radii
close to the central star, this is particularly evident for the 2:1
commensurability, where $R_i \sim 2 R_\oplus$ for $P_i < 10$ days, in
contrast to larger distances where larger planets abound. The decrease
in $R_i$ closer to the star, where the observed resonance offset is
larger, appears contrary to expectations from our model (\ref{eq14}),
which indicates that $\Delta _{(p+q)/p} \propto m_i/m_0$. As we see
later on, this problem can be partially overcome considering tidal
evolution during the lifespan of the system.

Finally, since our model requires estimations of the masses, the
bottom panel shows the mass-radius distribution of all known planets
(resonant or otherwise) whose values have errors below $40\%$. Two
different empirical fits are usually employed for this relation. The
$R_i^{2.06}$ power law from Lissauer et al. (2011) is shown as a
broken orange curve and was deduced from the values of Earth and
Saturn. As can be seen from the figure, it shows a fair fit for bodies
with $R_i > 5 R_\oplus$, but underestimates the mass of exoplanets
with smaller radii. Conversely, the fit presented by Weiss \& Marcy
(2014) shows a better agreement for smaller bodies, but still falls
short of the average values even in this domain. Many small planets have
been confirmed since Weiss \& Marcy (2014) and a similar analysis
performed today would lead to a different set of numerical values.

To obtain an updated empirical relation, we then searched for two best
fits, one for $R<R_{\rm crit}$ and one for larger bodies, where the
value of $R_{\rm crit}$ was chosen close to $4 R_{\oplus}$ but with
some slight freedom to reduce the overall variance. The result is
shown as two thick green lines. Since we also need to model the
dispersion around the mean values, the light green shaded areas
corresponds to the regions within $1\sigma_{\rm m}$.  Explicitly, we
obtained
\begin{equation}
\log{\biggl( \frac{m_i}{m_\oplus} \biggr)} \simeq 
\left\{
\begin{array}{ll}
\;\; 0.60 + 0.92 \, \log{\biggl( \frac{R_i}{R_\oplus} \biggr)} \;\;\; {\rm if} 
\;\; R \le 3.8 R_\oplus \\
-0.13 + 2.18 \, \log{\biggl( \frac{R_i}{R_\oplus} \biggr)} \;\;\; {\rm if} 
\;\; R > 3.8 R_\oplus \\
\end{array}
\right.
\label{eq18}
\end{equation}
plus a normal distribution centered around this value with dispersion
$\sigma_{\rm m} \simeq 0.42$. Both slopes are very similar to those
proposed by \cite{2014ApJ...783L...6W} and
\cite{2011ApJS..197....8L}. The value of $R_{\rm crit}$ was fine-tuned
to obtain a smooth transition between both segments.  Equation
(\ref{eq18}) henceforth allows us to perform Monte Carlo simulations
assigning masses to different observed systems with given values of
$R_i$.

To apply our model statistically we still require knowledge of the
disk parameters. Probable values are also weakly constrained. Flare
index are expected to be strongly dependent on heating and cooling
efficiencies and radiative properties of the disk, and
hydro-simulations have registered values on the order of $f \sim 0.3$
(e.g., \citealp{2013A&A...549A.124B}). Similarly, although classical
estimates for the minimum mass solar nebula (MMSN) point towards
values of $\alpha \sim 3/2$, there is no obvious reason to believe
such a value should be universal. The observed distribution of
exoplanetary systems shows a wide diversity of power-law index (e.g.,
\citealp{2014MNRAS.440L..11R}), although the present location may not
be indicative of their formation sites if planetary migration was wide
spread.  Additionally, the scale height $H_0$ is believed to be $H_0
\in [0.01,0.1]$ and most studies assume values $H_0 \simeq
0.05$. However, since the resonance offset is a strong function of
$H_0$ (in fact, $\Delta_{(p+q),p} \propto 1/H_0$, see equation
(\ref{eq14})), we kept its value as a free parameter.

\begin{figure*}[t!]
\centering
\includegraphics[width=0.98\textwidth,clip=true]{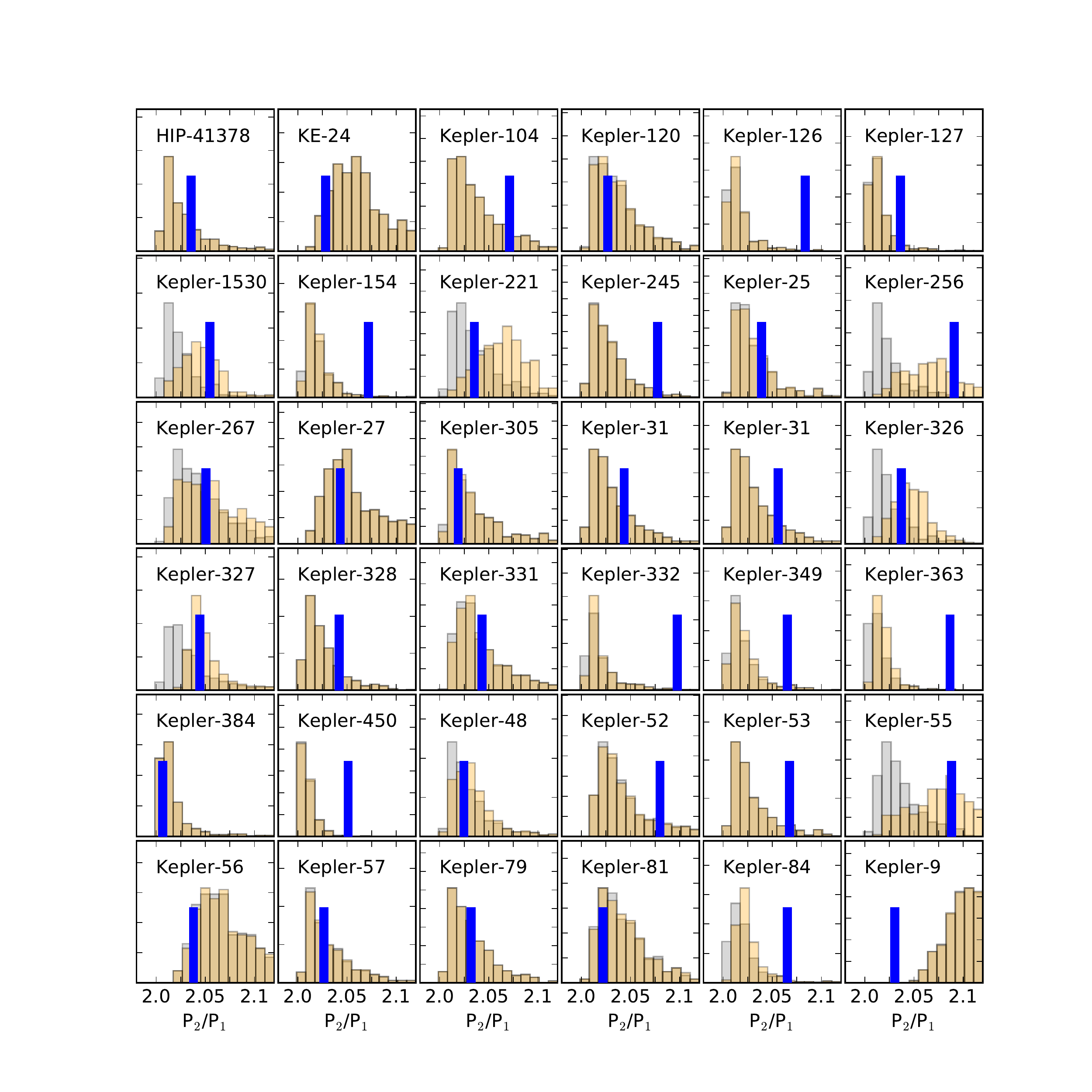} 
\caption{Blue columns indicate the observed orbital period ratio
  $P_2/P_1$ for 36 exoplanetary pairs in the vicinity of the 2:1
  resonance. The gray histograms are the result of 400 Monte Carlo
  simulations as obtained from equation (\ref{eq3}) with random
  assigned values for masses and disk parameters (see text for
  details). Orange histograms represent previous values evolved
  through tidal effects for the duration of the lifespans of the
  systems.}
\label{fig6}
\end{figure*}

\begin{figure*}[t!]
\centering
\includegraphics[width=0.98\textwidth,clip=true]{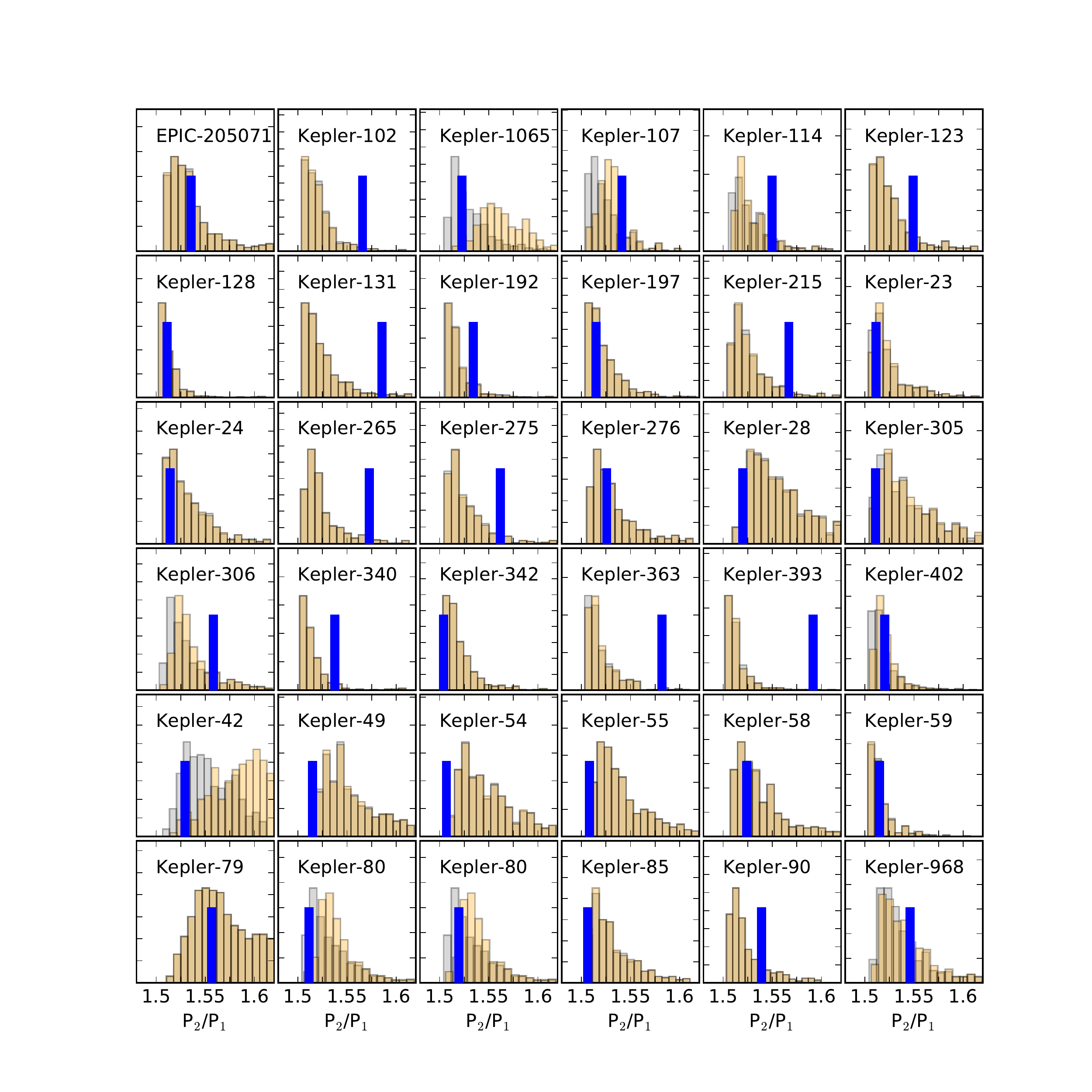} 
\caption{Same as previous figure but for systems in the vicinity of
  the 3:2 resonance.}
\label{fig7}
\end{figure*}

Our Monte Carlo simulations were performed as follows. For each
exoplanetary system in the vicinity of the 2:1 and 3:2 MMRs, we
assigned each planet a physical radius $R_i$ in accordance to the
estimated values and errors given in the data set. The masses were
then generated from expression (\ref{eq18}) including a normal
distribution with variance $\sigma_{\rm m}$. In each run the disk
properties $(f,\alpha,H_0)$ were also chosen randomly assuming normal
distributions with mean $(f_{\rm med},\alpha_{\rm med},{H_0}_{\rm
  med})$ and variances $(\sigma_{\rm f},\sigma_{\alpha},\sigma_{\rm
  {H_0}})$. The resulting parameters were then fed into equation
(\ref{eq14}) to obtain a value of the resonance offset. If the set of
parameters did not lead to convergent migration and a stable resonance
configuration (as described by the conditions deduced in
\citealp{2015ApJ...810..119D}), we repeated the process. We ran a
total of 400 successful cases were run for each system and plotted the
distribution of $P_2/P_1$ as histograms.

Figure \ref{fig6} shows results for 36 systems in the vicinity of the
2:1 resonance. Blue columns indicate the observed value of the orbital
period ratio, while the gray histograms correspond to our Monte Carlo
simulations with $f = 0.25 \pm 0.05$, $\alpha = 1.0 \pm 0.5$ and $H_0
= 0.03 \pm 0.02$.  The mean values were chosen close to those
estimated from Figure \ref{fig3}, except for the disk scale height
taken was slightly lower. The variances were chosen so as to include
the values usually present in the literature (e.g., $H_0 = 0.05$ and
$\alpha = 3/2$ for the MMSN). Although these values may be considered
arbitrary, they are consistent with current knowledge of
protoplanetary disks.

Although the results look promising, some of the larger offsets, such
as those observed for Kepler-126, Kepler-154, and Kepler-55, appear as
low-probability events. However, we must also allow for resonance
divergence through tidal evolution. We adopted the analytical
prescription described in \cite{2011CeMDA.111...83P} , which includes
tidal distortions in both the planets and central star. However, this
introduces additional free parameters into our model. Without any
information on their internal structure or chemical composition, it is
not possible to give reliable values for the planetary tidal
coefficient $Q_p$. Even within our own solar system these parameters
are still under discussion (e.g., \citealp{2016CeMDA.126..145L}),
although some order of magnitude is generally agreed upon. Since we
wanted to avoid fixing a single value for $Q_p$, we introduced the
following empirical formula:
\begin{equation}
\log{(Q_p)} = {\rm max}(2,6-0.7 \rho_p)
\label{eq19}
,\end{equation} where $\rho_p$ is the volumetric density of the
planet\ in units of gr/cm$^3$.  Even though this relation is not based
on any physical evidence, it yields qualitatively correct results in
the case of the Earth ($Q_p \sim 10^2$) and Jupiter ($Q_p \sim 10^5$);
it also gives a smooth transition between both extremes.

Following \cite{2011A&A...528A...2B} we adopted a stellar tidal
coefficient $Q_*=10^6$, while the stellar rotation was taken as $P_* =
15 \pm 5$ days.  However, stellar tides proved to be negligible in all
cases and the results were found to be virtually insensitive to both
$Q_*$ and $P_*$. Finally, few systems have estimated ages $T_{\rm
  age}$, so we assumed random values following an uniform distribution
in the interval $T_{\rm age} \in [2.5,7.5]$ Gyrs. All planets were
initially placed in spin-orbit equilibrium (pseudo-synchronous
rotation) regardless of their distance from the star.

New final values for the offsets, incorporating tidal evolution, are
shown as orange histograms in Figure \ref{fig6}. When tidal effects
are negligible, the orange values appear to be superimposed on the
original gray histograms (e.g., HIP-41378, KE-24). In other systems,
such as Kepler-256, Kepler-327, or Kepler-55, tidal evolution causes a
significant increase in offset leading to final values that are much
closer to those currently observed. Other systems, however, continue
to be difficult to explain by this mechanism (e.g., Kepler-126,
Kepler-154, Kepler-332, and Kepler-363).  Most of these are
characterized by very large offsets on the order of $\Delta_{2/1}
\simeq 0.08$, and could in fact constitute part of the background
secular population, which appear overlaid to the (near)-resonant
systems.

Three other problematic systems are worth mentioning. Kepler-450 has a
ratio $R_2/R_1 \simeq 3$ leading in most Monte Carlo runs to $m_2/m_1
\simeq 10$. As seen from equation (\ref{eq14}), for such systems our
model inexorably predicts very small offsets, much smaller than the
observed value of $\Delta_{2/1} \simeq 0.04$. An opposite result
perceived for Kepler-9, where our model always seems to predict larger
offsets than observed. This is not surprising since the bodies in this
system have assigned masses in the giant-planet range and should have
undergone Type-II migration. The inability of our model to predict
compatible values is thus expected. Another complicated system is
Kepler-84, which has a $\sim 48\%$ error in the determination of the
stellar radius; this error can lead to a huge mistake in the process
of measuring planetary radii.

In general terms, however, our model appears to work well. For most
real systems it predicts offsets similar to the observed values, even
if the agreement does not occur for the peak of the simulated
distribution. We do not believe this is an important issue since we do
not know the actual disk parameters that housed each individual
planets. Consequently, the fact that we are able to model the
diversity and approximate magnitude of $\Delta_{2/1}$ for so many real
cases is, in our opinion, a very positive result. It is important to
mention that, as proposed by \cite{2015MNRAS.453.4089S}, tidal effects
cannot by themselves explain the (near)-resonant
configurations. However, they are an important additional mechanism,
at least in some planets.

Figure \ref{fig7} shows results for 36 systems in the vicinity of the
3:2 mean-motion resonance. As before, our model underestimates the
offset for a few systems (e.g., Kepler-102, Kepler-131, Kepler-363,
and Kepler-393), but again these could be background objects whose
position in the vicinity of the resonant domain is merely
coincidental. The vast majority of the cases, however, are well
reproduced by the model and in several examples the observed value
closely matches the peak of the histograms.

As an additional test, we analyzed the sensitivity of these results
with respect to the model parameters. We found that the distribution
of the offsets was only weakly dependent on the disk flare and surface
density profile, although they did show a stronger relation with both
$H_0$ and $Q_e$. Higher values of both these quantities led to lower
$\Delta_{(p+q)/p}$, so more precise estimations in the future could
help further evaluate the merits of this approach.

Finally, although in this paper we concentrated on the (near)-resonant
population of exoplanets around the 2:1 and 3:2 MMRs, it is important
to keep in mind that most of the close-in systems lie in secular
configurations. While explaining the origin of such systems lies
beyond the scope of the present work, it is possible that at least
some of them could be the result of failed resonance captures or the
outcome of collisional or planet-scattering events. Not all mass
ratios or disk parameters necessarily lead to stable resonance
trapping (e.g., \citealp{2015ApJ...810..119D}), and the consequent
chaotic reshuffling of the orbits could explain at least some of our
(relatively few) negative results.

\section{Conclusions}

We have presented a very simple model for the resonance offset
$\Delta_{(p+1)/p}$ generated by Type-I migration in a flared disk and
applicable to first-order resonances. We described planetary migration
using analytical prescriptions with correcting factors that were
fine-tuned with hydrodynamical simulations of two-planet resonant
systems. We also included the post-formation orbital evolution from
tidal effects, both in the star and planetary bodies.

We found that the observed distribution of exoplanetary systems in the
vicinity of the 2:1 resonance presents compelling evidence of an
increase in the value of $\Delta_{2/1}$ closer to the central
star. Planetary pairs near the 3:2 resonance show much weaker
dependence and much smaller offsets for all distances. Our model
adequately reproduces both these properties and the general
distribution of the (near)-resonant {\it Kepler} systems is consistent
with smooth migration in a thin laminar flared disk with significant
flare (perhaps up to $f \sim 0.25$) and very thin disks ($H_0 \sim
0.03$).

Validation of this model proved difficult since most observed systems
lack reliable estimations of the planetary masses. However, in most
cases Monte Carlo simulations with an updated empirical mass-ratio
relation and random values for the disk parameters, led to
distributions of the resonance offsets similar to the observed
values. Although some cases were not adequately explained, indicating
that our model is still insufficient, we were able to reproduce values
of $\Delta_{(p+1)/p}$ that were at least qualitatively similar and, in
some cases, very close to the detected magnitudes.

Finally we found that large offsets, even on the order of $\Delta_{2/1}
\simeq 0.1$, are still associated with libration of one or both
critical arguments.  Without a more sophisticated resonance model we
cannot say whether this libration is dynamical (i.e., inside the
separatrix) or kinematic (i.e., outside the main resonance domain);
however, it is possible that several of these (near)-resonant systems
may be resonant after all.

\begin{acknowledgements}
This work has been supported by research grants from ANCyT, CONICET,
and Secyt-UNC. The authors are grateful to an anonymous referee for
inspiring questions that helped improve the work. We also wish to
thank IATE and CCAD (Universidad Nacional de C\'ordoba) for extensive
use of their computational facilities.
 
\end{acknowledgements}

\bibliographystyle{aa}
\bibliography{biblio}

\end{document}